\documentclass[a4paper]{article}

\usepackage{INTERSPEECH2018}
\usepackage{amsmath}
\usepackage{hyperref}
\usepackage{multirow}
\usepackage{xr}
\usepackage{hhline}
\usepackage{algorithm}

\newcommand*{\enc}{\operatorname{enc}}
\newcommand*{\dec}{\operatorname{dec}}
\newcommand*{\gen}{\operatorname{gen}}

\newcommand*{\emb}{\operatorname{emb}}
\newcommand*{\dis}{\operatorname{D}}
\newcommand*{\UNI}{\operatorname{UNIFORM}}
\newcommand*{\X}{\mathcal{X}}
\newcommand*{\Y}{\mathcal{Y}}
\newcommand*{\D}{\mathcal{D}}

\title{Multi-target Voice Conversion without Parallel Data by Adversarially Learning Disentangled Audio Representations}
\name{Ju-chieh Chou$^1$, Cheng-chieh Yeh$^1$, Hung-yi Lee$^1$, Lin-shan Lee$^1$}
\address{
  $^1$College of Electrical Engineering and Computer Science, National Taiwan University}
\email{\{r06922020,r06942067,hungyilee\}@ntu.edu.tw,lslee@gate.sinica.edu.tw}

\begin{document}

\maketitle
\begin{abstract}
Recently, cycle-consistent adversarial network (Cycle-GAN) has been successfully applied to voice conversion to a different speaker without parallel data, although in those approaches an individual model is needed for each target speaker. In this paper, we propose an adversarial learning framework for voice conversion, with which a single model can be trained to convert the voice to many different speakers, all without parallel data, by separating the speaker characteristics from the linguistic content in speech signals. An autoencoder is first trained to extract speaker-independent latent representations and speaker embedding separately using another auxiliary speaker classifier to regularize the latent representation. The decoder then takes the speaker-independent latent representation and the target speaker embedding as the input to generate the voice of the target speaker with the linguistic content of the source utterance. The quality of decoder output is further improved by patching with the residual signal produced by another pair of generator and discriminator. A target speaker set size of 20 was tested in the preliminary experiments, and very good voice quality was obtained. Conventional voice conversion metrics are reported. We also show that the speaker information has been properly reduced from the latent representations.

\end{abstract}

\noindent\textbf{Index Terms}: voice conversion, disentangled representation, adversarial training.

\section{Introduction}

Speech signals inherently carry both linguistic and acoustic information. Voice conversion (VC) aims to convert the speech signals from a certain acoustic domain to another while keeping the linguistic content unchanged. Examples of acoustic domains may include speaker identity~\cite{stylianou1998continuous, kain1998spectral, miyoshi2017voice, nakashika2014voice, saito2011one, kinnunen2017non}, speaking style, accent, emotion~\cite{kaneko2017sequence} or some other properties orthogonal to the linguistic content. Voice conversion (VC) can be used for various tasks such as speech enhancement~\cite{toda2012statistical,kumar2016improving}, language learning for non-native speakers~\cite{kaneko2017sequence}, to name a few~\cite{inanoglu2009data}. This work focuses on the conversion of speaker identity.

In general, among the difficult problems for VC approaches, the need of aligned data~\cite{desai2010spectral,mohammadi2014voice,nakashika2014high,sun2015voice, chen2014voice, takamichi2014postfilter, ohtani2006maximum}, and over-smoothing of signals~\cite{stylianou1998continuous,toda2007voice,helander2010voice} are two examples carefully studied. Due to the difficulties in obtaining aligned corpora, approaches utilizing generative models such as Variational Autoencoders (VAEs)~\cite{kingma2013auto} and Generative Adversarial Networks (GANs)~\cite{goodfellow2014generative} were studied because they can be trained with non-parallel data. With VAEs, the encoder learns the speaker-independent linguistic content, which can then be used by the decoder to generate the voice with a specified speaker id~\cite{hsu2016voice,hsu2017voice,hsu2017unsupervised}. Cycle-consistent adversarial network (Cycle-GAN) was also used to learn the mapping from the source speaker to the target speaker in an unsupervised way~\cite{kaneko2017parallel,gao2018voice}.

Some prior works successfully used VAEs for VC, but generated the voice frame-by-frame~\cite{hsu2017voice}. Some other prior works were able to disentangle the linguistic content from the speaker characteristics when learning the representations, but based on some heuristic assumptions~\cite{hsu2017unsupervised}. 
Cycle-GAN was used for VC without parallel data, but an individual model is needed for each target speaker~\cite{kaneko2017parallel,gao2018voice}. In this paper, we propose an autoencoder architecture which is able to deal with several frames at a time, leading to better results because in this way information carried by neighboring frames can be considered. This approach also uses jointly trained speaker classifier to remove the need for heuristic assumptions made previously. This approach is able to train a single model to convert the voice to many different speakers, all without parallel data, by separating the speaker characteristics from the linguistic content. This is similar to some degree to some works in computer vision which learned disentangled representation~\cite{lample2017fader}, or shared generator with conditional input~\cite{choi2017stargan}.

\begin{figure}[t]
  \centering
  \includegraphics[width=\linewidth]{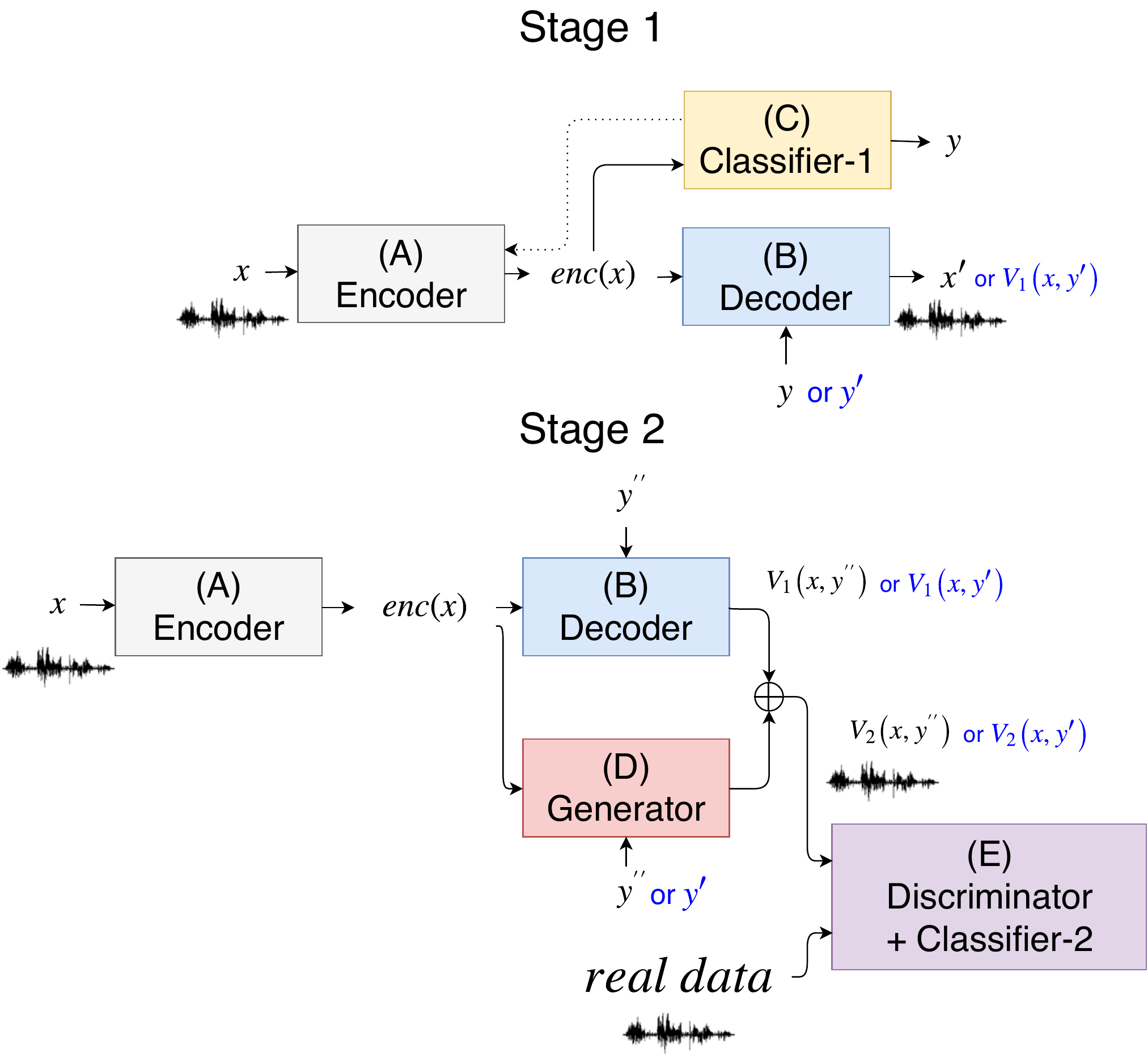}
  \caption{The training procedure. In stage 1, the encoder (Block(A)) is trained to generate speaker-independent representation with the regularization of the classifier-1 (Block(C)). The decoder (Block(B)) is trained to reconstruct the original acoustic features. In stage 2, a generator (Block(D)) is trained to enhance the output of the decoder (Block(B)) with the help of the Discriminator plus Classifier-2 (Block(E)).}
  \label{fig:training}
\end{figure}

The proposed approach includes two stages of training as in \autoref{fig:training}
\footnote{The $\oplus$ in~\autoref{fig:training} indicates element-wise addition.}. 
In stage 1, we train an autoencoder. The encoder encodes the input spectra into a latent representation for the linguistic content but without speaker characteristics based on the adversarial training concept. This is achieved by training a classifier to classify the speaker based on the latent representation, while the encoder is trained adversarially to fool the classifier. On the other hand, the decoder merges the speaker identity with the latent representation to reconstruct the original spectra. 

In stage 2, we train another generator to generate the residual signal (or fine structure) of the decoder output. The decoder output is patched with the residual signal to be the final output. The generator is in turn learned with a discriminator which outputs a scalar to indicate whether the input signal is realistic. The discriminator is further trained with an auxiliary classifier, predicting the speaker for the input signal. This helps the generator to produce signals carrying more characteristics of the target speaker.

\section{Proposed approach}


Let $x \in \X$ be an acoustic feature sequence where $\X$ is the collection of all such sequences, and $y \in \Y$ be a speaker where $\Y$ is the group of all speakers who produce the sequence collection $\X$. The training set $\D=\{(x_1, y_1)\dots(x_m, y_m)\}$ contains m pairs of $(x_i, y_i) \in (\X, \Y)$, where the sequence $x_i$ is produced by speaker $y_i$. During training, $x$ is a fixed-length segment randomly sampled from $\X$. During testing, $x$ can have variable length because the model here is built with recurrent-based components. The whole framework includes two stages of training as shown in~\autoref{fig:training} and explained below.

\subsection{Stage 1: Autoencoder plus classifier-1}

This stage is to learn an autoencoder plus classifier-1.

\newpage
\textbf{Autoencoder:}

The encoder (Block(A)) is trained to map an input sequence $x$ to a latent representation $\enc(x)$. The decoder (Block(B)) is trained to generate $x'$ which is a reconstruction of $x$ from $\enc(x)$ given the speaker identity $y$. 



\begin{equation}
        x'=\dec(\enc(x), y).
\end{equation}
The Mean Absolute Error (MAE) is minimized in training the autoencoder because this generates sharper output than mean square error~\cite{isola2017image}. So the reconstruction loss is given as in~\eqref{eq1}
\begin{equation}
        L_{rec}(\theta_{enc}, \theta_{dec})= \sum_{(x, y) \in \D} \Arrowvert x'-x \Arrowvert_1 ,
  \label{eq1}
\end{equation}
where $\theta_{enc}$ and $\theta_{dec}$ are the parameters of the encoder and decoder respectively. This autoencoder alone can achieve VC as below. Given an utterance $x$ produced by a source speaker $y$, the decoder can generate the voice of a target speaker $y'$ using the linguistic content of $x$,
\begin{equation}
        V_1(x,y') =\dec(\enc(x), y').
\end{equation}
During training, the decoder input is $\enc(x)$ and  $y$, but during voice conversion, we replace $y$ with the target speaker $y'$. $V_1(x,y')$ is the output of stage 1, which has the linguistic content of $x$, but the identity of $y'$. 

\textbf{Classifier-1:} 
The autoencoder itself learned with~\eqref{eq1} cannot make the latent representation $\enc(x)$ speaker-independent. The speaker characteristics of the original speaker $x$ existing in $\enc(x)$ inevitably degrades the performance of VC. This is why we train classifier-1 (Block(C)) in addition to regularize the autoencoder to make $\enc(x)$ speaker-independent. For each training pair $(x_i, y_i) \in \D$, the classifier takes $\enc(x_i)$ as input and outputs the probability $P_{cls1}(y|\enc(x_i)), y \in \Y$, which is the probability that $x_i$ is produced by speaker $y$. The classifier-1 is trained to minimize the negative log-probability to differentiate the different speakers, as in~\eqref{eq2},
\begin{equation}
    \begin{aligned}
        L_{cls1}(\theta_{enc}, \theta_{cls1})= \sum_{(x_i, y_i) \in \D} - \log P_{cls1}(y_i|\enc(x_i)).
    \end{aligned}
  \label{eq2}
\end{equation}
On the other hand, however, the encoder is trained to maximize~\eqref{eq2} in order to remove the speaker identity in $\enc(x)$. So the full objective for the autoencoder regularized by clasifier-1 is 
\begin{equation}
    \begin{aligned}
        L_{ae}(\theta_{enc}, \theta_{dec}, \theta_{cls1}) = L_{rec}(\theta_{enc}, \theta_{dec}) - \lambda  L_{cls1}(\theta_{enc}, \theta_{cls1}),
    \end{aligned}
  \label{eq4}
\end{equation}
which integrates~~\eqref{eq1} and~\eqref{eq2} and $\lambda$ is a hyper-parameter. The autoencoder and the classifier are trained alternatively. 
%
%


\subsection{Stage 2: GAN}

If we simply perform VC with stage 1 as mentioned above, even with the help of classifier-1, the reconstruction loss tends to generate blurry spectra and artifact. This is why in stage 2, we train another pair of generator and discriminator to guide the output spectra to be more realistic~\cite{isola2017image}.

Based on the concept of decoupled learning~\cite{zhang2018decoupled}, which stabilize GAN training by decoupling decoder and generator, we separately train another generator (Block(D)) taking $enc(x)$ and the speaker identity $y''$, which is a uniformly sampled speaker out of all speakers in $\Y$, as the input and generate the residual (or fine structure of the signal) of the output of the decoder (Block(B)). The parameters of the encoder (Block(A)) and the decoder (Block(B)) are fixed in this training stage, which stabilizes the training procedure.

As shown in~\autoref{fig:training}, the generator here is trained with the help of a "discriminator plus classifier-2" (Block(E)), and the final output during VC test for a selected target speaker $y' \in \Y$ is the addition of the output of the decoder (Block(B)) and the generator (Block(D)), or $y'' = y'$ in~\autoref{fig:training},
\begin{equation}
V_2(x,y') = V_1(x,y') + \gen(\enc(x), y').
\label{eq6}
\end{equation}
In~\eqref{eq6}, $x$ is the input speech, $\enc(x)$ the encoder output, $y'$ the selected target speaker, $V_1(x,y')$ is the converted voice obtained in the stage 1, $\gen(\enc(x), y')$ the output of the generator, and $V_2(x,y')$ is the VC result for stage 2.

The generator is learned with a discriminator (in Block(E)) in an adversarial network. This discriminator is trained to distinguish whether an input acoustic feature sequence, $x$, is real or generated by machine. The output of the discriminator $\dis(x)$ is a scalar indicating how real $x$ is, the larger $\dis(x)$, the more possible $x$ is real. This adversarial network is trained with the loss in~\eqref{eq7}, where $\theta_{gen}$ and $\theta_{dis}$ are the parameters for the generator and the discriminator.
\begin{equation}
    \begin{aligned}
        &L_{adv}(\theta_{gen}, \theta_{dis})=\sum_{x \in \D} \log(\dis(x)) +\\
        & \qquad \sum_{x \in \D, y''\sim \UNI(\Y)}  \log(1-\dis(V_2(x,y''))).
    \end{aligned}
  \label{eq7}
\end{equation}
The discriminator gives larger values to real speech $x$ from the dataset $\D$ in the first term on the right of~\eqref{eq7}, while assigns lower score to the converted speech $V_2(x,y'')$ in the second term, where $y''$ is a speaker sampled uniformly from $\Y$. So the discriminator is trained to distinguish real voice and the generated data by maximizing $L_{adv}$ in~\eqref{eq7}, while on the other hand the generator is trained to fool the discriminator by minimizing $L_{adv}$ in~\eqref{eq7}.

In addition, the "discriminator plus classifier-2" also includes a classifier-2 which learns to predict the speaker for the speech $x$ by generating a distribution of speakers $P_{cls2}(y|x)$~\cite{odena2016conditional} based on the training data $\D$. This classifier-2 is trained by minimizing the loss in~\eqref{eq8},
\begin{equation}
    \begin{aligned}
        &L^d_{cls2}(\theta_{dis})=\sum_{(x_i,y_i) \in \D} - \log P_{cls2}(y_i|x_i).
    \end{aligned}
  \label{eq8}
\end{equation}
This classifier-2 and the discriminator share all layers except with separated last layer. On the other hand, the generator should learn to generate the voice $V_2(x,y'')$ for a uniformly sampled speaker $y'' \in \Y$ which can be predicted as the voice of $y''$ by the classifier-2, which implies $V_2(x,y'')$ preserves more speaker characteristics of $y''$. So the generator should be trained to minimize the loss in \eqref{eq9},
\begin{equation}
    \begin{aligned}
        &L^g_{cls2}(\theta_{gen})=\sum_{x \in \D, y''\sim \UNI(\Y)}  
        -\log P_{cls2}(y''|V_2(x,y'')).
    \end{aligned}
  \label{eq9}
\end{equation}
Here~\eqref{eq8} and~\eqref{eq9} are exactly the same, except~\eqref{eq8} is for real data and correct speakers in the data set, while~\eqref{eq9} for generated voice and sampled target speakers.

So the complete loss function in stage 2 is given as~\eqref{eq10} and~\eqref{eq11} respectively for the generator $\theta_{gen}$ and the discriminator $\theta_{dis}$, trained alternatively, where the first terms on the right of~\eqref{eq10}~\eqref{eq11} are in~\eqref{eq7}, and the second terms of~\eqref{eq10}~\eqref{eq11} are in~\eqref{eq8}~\eqref{eq9}.
\begin{equation}
    \begin{aligned}
        &L_{gen}(\theta_{gen})=L_{adv}(\theta_{gen}, \theta_{dis})+L^g_{cls2}(\theta_{gen})
    \end{aligned}
  \label{eq10}
\end{equation}
\begin{equation}
    \begin{aligned}
        &L_{dis}(\theta_{dis})=-L_{adv}(\theta_{gen}, \theta_{dis})+L^d_{cls2}(\theta_{dis})
    \end{aligned}
  \label{eq11}
\end{equation}
%

\begin{table}[ht!]
    \centering
    \caption{Network architectures. C indicates convolution layer. FC indicates fully-connected layer. Conv1d-bank-K indicates convolution layer with kernel size from 1 to K. LReLU indicates leakyReLU activation. IN indicates instance normalization~\cite{DBLP:journals/corr/UlyanovVL16}. Res indicates residual connection. PS indicates pixel shuffle layer for upsampling. The kernel size K for discriminator is 64-128-256-512-512.}
    \label{tab:network}
    \begin{tabular}{ll}
        \hline
        \multicolumn{2}{c}{\textbf{Encoder}}                                                                                                                \\ \hline \hline
        \multicolumn{1}{l|}{conv-bank block}        & Conv1d-bank-8, LReLU, IN                                                                              \\ \hline
        \multicolumn{1}{l|}{conv block $\times$ 3}  & \begin{tabular}[c]{@{}l@{}}C-512-5, LReLU\\ C-512-5, stride=2, LReLU, IN, Res\end{tabular}            \\ \hline
        \multicolumn{1}{l|}{dense block $\times$ 4} & FC-512, IN, Res                                                                                       \\ \hline
        \multicolumn{1}{l|}{recurrent layer}        & bi-directional GRU-512                                                                                \\ \hline
        \multicolumn{1}{l|}{combine layer}          & recurrent output + dense output                                                                       \\ \hline
                                                    &                                                                                                       \\ \hline\hline
        \multicolumn{2}{c}{\textbf{Decoder/Generator}}                                                                                                                \\ \hline\hline
        \multicolumn{1}{l|}{conv block $\times$ 3}  & \begin{tabular}[c]{@{}l@{}}$\emb_l(y)$, C-1024-3, LReLU, PS\\ C-512-3, LReLU, IN, Res\end{tabular} \\ \hline
        \multicolumn{1}{l|}{dense block $\times$ 4} & $\emb_l(y)$, FC-512, IN, Res                                                                          \\ \hline
        \multicolumn{1}{l|}{recurrent layer}        & $\emb_l(y)$, bi-directional GRU-256                                                                   \\ \hline
        \multicolumn{1}{l|}{combine layer}          & recurrent output + dense output                                                                       \\ \hline
                                                    &                                                                                                       \\ \hline\hline
        \multicolumn{2}{c}{\textbf{Classifier-1}}                                                                                                             \\ \hline\hline
        \multicolumn{1}{l|}{conv block $\times$ 4}  & \begin{tabular}[c]{@{}l@{}}C-512-5, LReLU\\ C-512-5, IN, Res\end{tabular}                                 \\ \hline
        \multicolumn{1}{l|}{softmax layer}          & FC-$N_{speaker}$                                                                                      \\ \hline
                                                    &                                                                                                       \\ \hline\hline
        \multicolumn{2}{c}{\textbf{Discriminator}}                                                                                                          \\ \hline\hline
        \multicolumn{1}{l|}{conv block $\times$ 5}  & C-K-5, stride=2, LReLU, IN                                                                            \\ \hline
        \multicolumn{1}{l|}{conv layer}             & C-32-1, LReLU, IN                                                                                                \\ \hline
        \multicolumn{1}{l|}{output layer}           & scalar output, FC-$N_{speaker}$(classifier-2) \\ \hline
    \end{tabular}
\end{table}

\begin{figure*}[t]
  \centering
  \includegraphics[width=\textwidth]{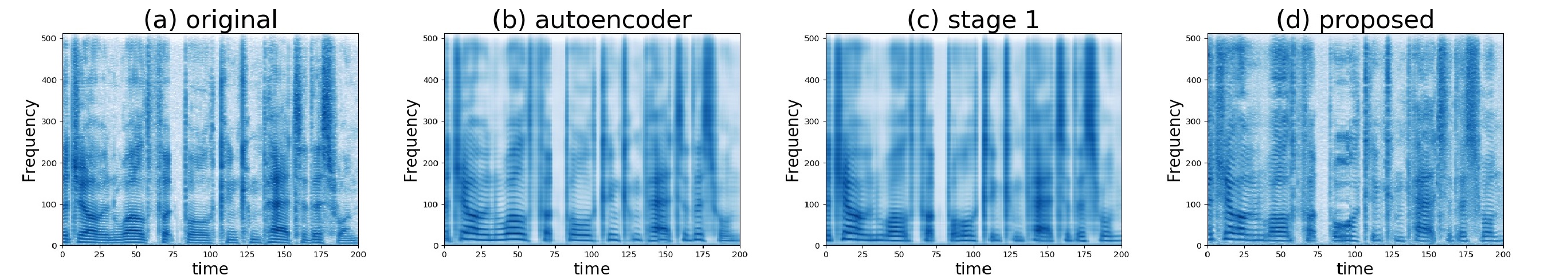}
  \caption{The heatmaps of the spectrogram: (a) original voice, (b)(c)(d) converted voice, (b) autoencoder of stage 1 alone without classifier-1, (c) complete Stage 1 with classifier-1, and (d) proposed approach incluiding Stage 1 and 2.}
  \label{fig:spec}
\end{figure*}
\begin{table}[ht!]
    \centering
    \caption{Spectral analysis/synthesis setting}
    \label{tab:feature}
    \begin{tabular}{l|l}
        pre-emphasis & 0.97        \\ \hline
        frame length & 50 ms       \\ \hline
        frame shift  & 12.5 ms     \\ \hline
        window type  & Hann        \\ \hline
        Sample rate  & 16kHz        \\ \hline
        Vocoder      & Griffin-Lim 
    \end{tabular}
\end{table}

\section{Implementation}

We adopted the model architecture from CBHG module~\cite{wang2017tacotron}. The detailed network architecture is listed in~\autoref{tab:network}. We did not use fully-connected layer across time-steps in order to deal with variable-length input. The convolution-bank aimed to capture local information about the acoustic features. We used the pixel shuffle layer to generate higher resolution spectra~\cite{shi2016real}. $\emb_l(y)$ indicates the speaker embedding in the l-th layer since the network may need different information in each layer. We plugged the embedding by adding it on the feature map. We used 1d convolution for every network except for the discriminator, which was built with 2d convolution to better capture the texture. 

\textbf{Dropout:} We provided the required noise in training with dropout in encoder as suggested~\cite{isola2017image}. We found it useful to add dropout in the classifier to improve the robustness of the model. We use 0.5 dropout rate in the encoder, and 0.3 in the classifier. 

\textbf{WGAN-GP:} GAN is notoriously hard to train. So we applied a different objective function, Wasserstein GAN with gradient penalty (WGAN-GP), to stabilize the training  process of GAN~\cite{gulrajani2017improved}.
%

\textbf{Hyper-parameters:} In training stage 1, if we add the classification loss $L_{cls1}$ in~\eqref{eq4} at the beginning of the training process, the autoencoder will have problems to reconstruct the spectra well. So we linearly increase the hyper-parameter $\lambda$ from 0 to 0.01 in the first 50000 mini-batches to make sure the latent representation became speaker-independent gradually. 
\footnote{Source code:
\url{https://github.com/jjery2243542/voice_conversion}}

\section{Experiments}

We evaluated our VC model on CSTR VCTK Corpus~\cite{veaux2017cstr}. The audio data were produced by 109 speakers in English with different accents, such as English, American, and India. Each speaker uttered different sets of sentences. We selected a subset of 20 speakers, 10 females and 10 males, as $\Y$ mentioned above. The dataset was randomly split to training and testing sets by the percentage 90\% and 10\%. 

We used log-magnitude spectrogram as the acoustic features. The detailed spectral analysis and synthesis setting was the same as the previous work~\cite{wang2017tacotron}. The detailed setting is in \autoref{tab:feature}.

\textbf{Training details:} We trained the network using Adam optimizer with learning rate $lr=0.0001$, $\beta_1=0.5, \beta_2=0.9$. Batch size was 32. We randomly sampled 128 frames of spectrogram with overlap. We trained classifier/discriminator for 5 iterations and 1 iteration for encoder/generator. 

We first pretrained the encoder and decoder with $L_{rec}$ in~\eqref{eq1} for $8000$ mini-batches, then pretrained the classifier-1 in with $L_{cls1}$ in ~\eqref{eq2} for $20000$ mini-batches. 
Stage 1 was trained for $80000$ mini-batches, and stage 2 another $50000$ mini-batches. 

\subsection{Objective evaluation}
\begin{figure}[t]
  \centering
  \includegraphics[width=\linewidth]{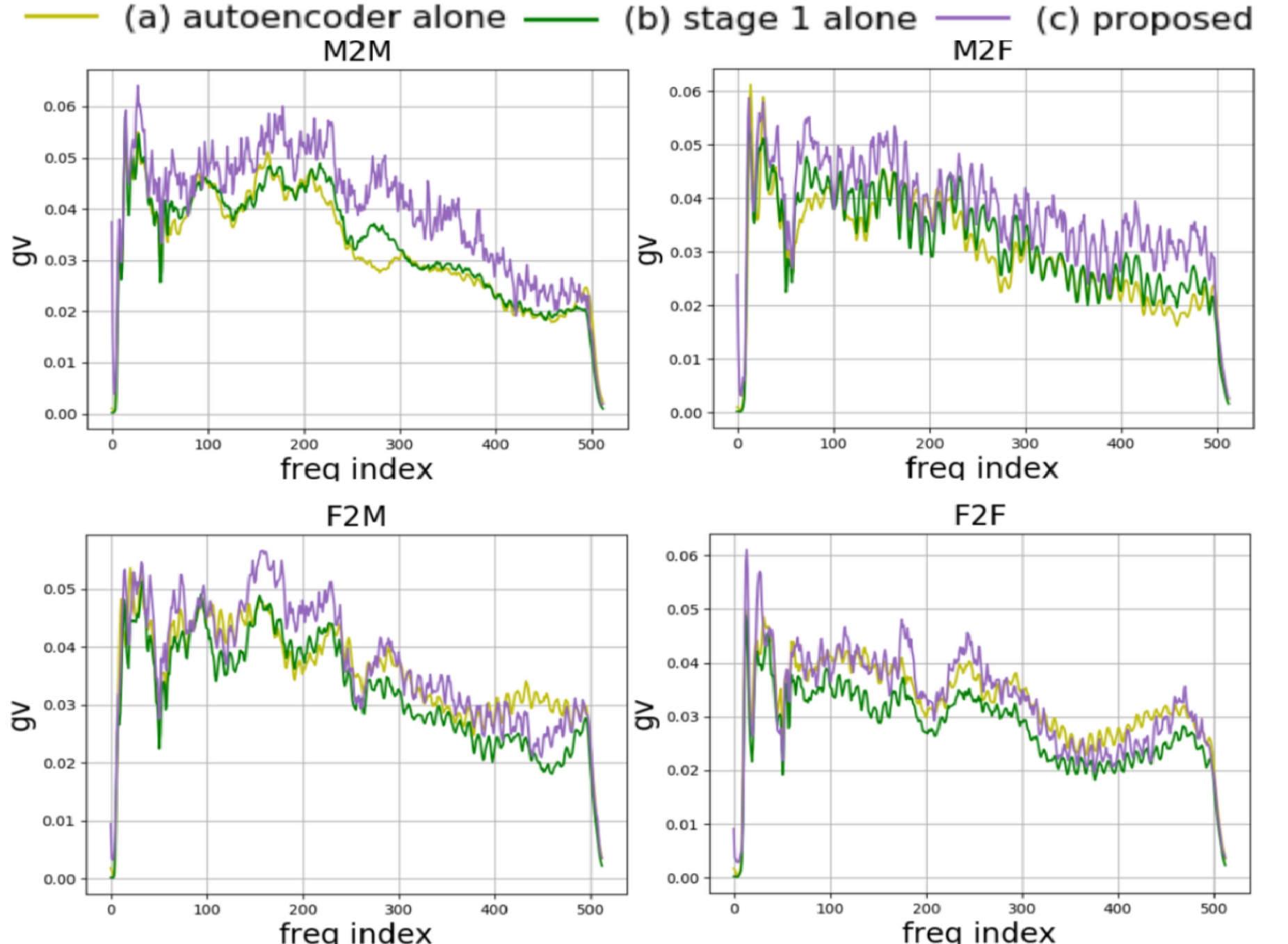}
  \caption{The global variance for each frequency index of spectrogram for 4 conversion examples: M2M, M2F, F2M, F2F, where F indicates female speaker and M indicates male speaker.}
  \label{fig:gv}
\end{figure}

Diversified distribution over all frequencies is a highly desired property of speech signals, and the over-smoothed spectra generated by many conventional approaches has been a major problem of voice conversion~\cite{toda2007voice}. This property can be observed by calculating the Global Variance (GV) over the spectrum. Higher global variance indicates sharpness of the converted speech. We evaluated the global variance for each of the frequency index for 4 conversion examples: male to male, male to female, female to male, and female to female. The results are in \autoref{fig:gv}. In each example 3 curves for 3 cases are plotted: (a) autoencoder of stage 1 alone without classifier-1, (b) complete stage 1 with classifier-1 and (c) proposed approach with stage 1 and stage 2. We can see in all cases the proposed approach (curves (c)) offered the best sharpness. Averages over all frequencies for those curves in~\autoref{fig:gv} are listed in~\autoref{tab:gv}, from which it is clear the proposed approach (row(c)) offered the highest global variance. A set of example spectrogram is in \autoref{fig:spec} for the original voice (a) and converted (b) (c) (d), where the sharpness offered by the proposed approach can be observed. 

\begin{table}[]
\centering
\caption{The averages of the global variance for 4 conversion examples: M2M, M2F, F2M, F2F, where F indicates female speaker and M indicates male.}
\label{tab:gv}
\begin{tabular}{l|llll}
                      & M2M    & \multicolumn{1}{c}{M2F} & \multicolumn{1}{c}{F2M} & \multicolumn{1}{c}{F2F} \\ \hline
(a) autoencoder alone & 0.0340 & 0.0334                  & 0.0341                  & 0.0307                  \\
(b) stage 1 alone     & 0.0326 & 0.0338                  & 0.0322                  & 0.0272                  \\
(c) proposed          & 0.0394 & 0.0401                  & 0.0389                  & 0.0333                 
\end{tabular}
\end{table}

\subsection{Subjective evaluation}

\begin{figure}[t]
  \centering
  \includegraphics[width=\linewidth]{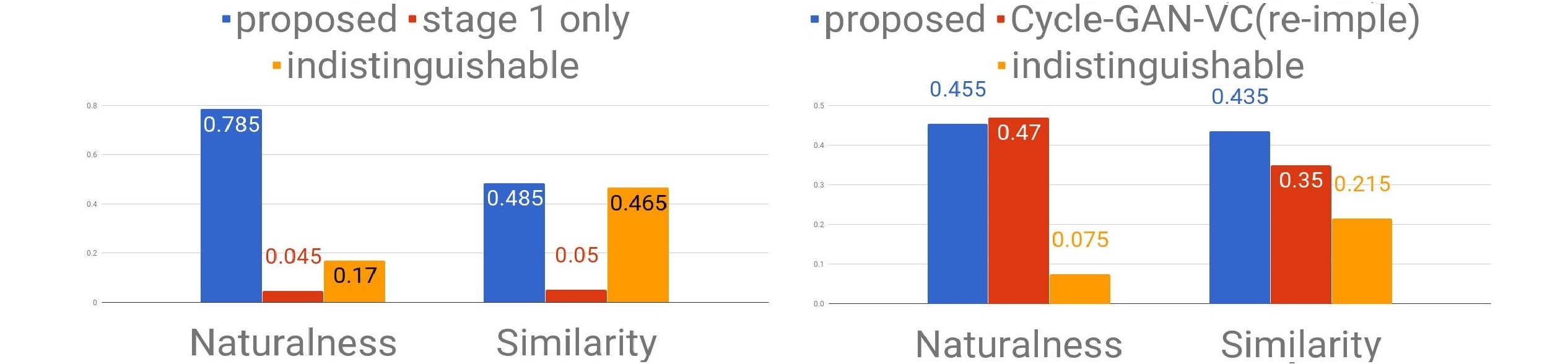}
  \caption{Results of subjective preference test in naturalness and similarity in speaker characteristics. The left compared the proposed approach with stage 1 only. The right compared the proposed approach with Cycle-GAN-VC~\cite{kaneko2017parallel}.}
  \label{fig:chart}
\end{figure}

We also performed subjective human evaluation for the converted voice. 20 subjects were given pairs of converted voice in random order and asked which one they preferred in terms of two measures: the naturalness and the similarity in speaker characteristics to a referenced target utterance produced by the target speaker.  
Average of examples including intra-gender and inter-gender conversion are shown in~\autoref{fig:chart}. The ablation experiment on the left of~\autoref{fig:chart} compared two method: the proposed approach including stages 1 and 2, and complete stage 1 with classfier-1 but not stage 2. We can see stage 2 has significantly improved the voice quality in terms of both naturalness and similarity in speaker characteristics. Here we also compared the proposed approach with a re-implementation of Cycle-GAN-VC~\cite{kaneko2017parallel}, a previous work comparable to methods utilizing parallel data. The result is on the right of~\autoref{fig:chart}. As the result shows, the proposed approach is comparable in terms of the naturalness and the similarity in speaker characteristics, while considering to multi-speakers without parallel data.
\footnote{Demo webpage:
\url{https://jjery2243542.github.io/voice_conversion_demo/}}


\subsection{Degree of disentanglement}
To evaluate the degree of disentanglement of our model with respect to speaker characteristics, we trained another speaker verification network that takes the latent representation $\enc(x)$ as input to predict the speaker identity~\cite{denton2017unsupervised}. The speaker verification network has the same architecture as the classifier-1 in stage 1. The verification accuracy was 0.916 without the classifier-1, but dropped significantly to 0.451 when classifier-1 was added. This verified that the classifier-1 successfully disentangled the speaker characteristics from the latent representation.



\section{Conclusion}
We proposed an approach for voice conversion by extracting the speaker-independent representation. No parallel data are needed and conversion to multiple target speakers can be achieved by a single model. We show that adding a residual signal can improve significantly the quality of converted speech. Objective evaluation metrics of global variance show that sharp voice spectra can be produced with this approach. This is also verified with subjective human evaluation.




\end{document}